\documentclass[12pt]{article}

    \usepackage[T1]{fontenc}
    \usepackage{graphicx,epsfig,amsmath,amsfonts,cite}
    \renewcommand{\abstract}{}
\begin{document}

\title{Hydrodynamic modes of conducting liquid in
random magnetic field}

\author{A. A. Stupka}
\maketitle
\begin{center} {\small\it Oles Honchar Dnipropetrovs'k National University, Gagarin
ave., 72, 49010 Dnipropetrovs'k, Ukraine\\antonstupka@mail.ru }
\end{center}


\begin{abstract}
Hydrodynamics of plasma in the random magnetic field is
considered, which is characterized by the second moment of
magnetic induction. Equations of ideal magnetic hydrodynamics in
such field are received for an adiabatic process. It is shown that
in Euler Equation a new variable is second moment of magnetic
induction enters, for which is received time Equation on the basis
of Maxwell Equations
 in the magnetohydrodynamic approximation. The one dimension plane
waves in this system are studied and the values of their phase
velocities are received.

\end{abstract}


\section*{Introduction}

In magnetohydrodynamics (MHD) an environment is considered as
single liquid and is described by the local values of density
$\rho $, pressure $P$
 and velocity $\vec v$. Except for these cleanly hydrodynamic values the state of
environment is characterized usual by magnetic induction $\vec B$
 (\cite{[2]} p. 8), which has some unzeroing value in the equilibrium state. We will suppose,
 that the magnetic field is a accidental value with the second moment of magnetic induction $\vec B$
different from a zero  \cite{[3]}, which is not taken to the
square of the first.
 Maxwell  tensor of tensions (\cite{[2]} p. 16) quadratic on the magnetic field is included in standard Euler Equation,
 for which we will build time Equation in MHD approximation.
 The offered generalization of MHD will allow to apply a theory to the systems
 in the accidental magnetic field, such as space plasma \cite{[4]} and plasma of solid.


\section{Equations of ideal MHD in the accidental field}

We will write out standard Equations of ideal MHD, ignoring all
dissipative effects (viscidity, heat conductivity and electric
resistance) (\cite{[2]} p. 19-24). Thus MHD effects show up most
brightly. Continuity equation is \begin{equation}\label{(1)}
\partial _t \rho  + div\rho \vec v = 0,
\end{equation}  Euler equation is
\begin{equation}\label{(2)}
d\rho \vec v/dt = - \nabla P - [\vec B,\vec j]/c,
\end{equation}  here
it is taken into account, that an environment unmagnetic and
strength of magnetic field coincides with induction $\vec H = \vec
B$. In addition, we will write down Maxwell Equations in MHD case
for magnetic induction as accidental value
\begin{equation}\label{(3)} rot\vec B = 4\pi \vec j/c,
\end{equation} \begin{equation}\label{(4)}
\partial _t \vec B = rot[\vec v,\vec B].
\end{equation}
At the construction of Eq. \ref{(4)} it is taken into account,
that at slow motions of environment electrons have time to be
displaced toward the increase electric potential so that gradient
of this potential will appeal to the zero. Thus the electric field
in the own frame of reference (\cite{[5]}  p.89) equal to  the
zero, i.e. $\vec E = - [\vec v,\vec B]/c$. We will put an electric
current from \ref{(3)} in \ref{(2)} and we will receives the law
of momentum saving  in a form
\begin{equation}\label{(5)}
\partial _t \left( {\rho v_i } \right) + \partial _k \pi _{ik}  =
0,
\end{equation} here denotations for derivative $\partial
/\partial x_k =
\partial _k $
 and for stream of momentum tensor $\pi _{ik}  = \rho v_i v_k  + P\delta _{ik}
  - \left( {B_i B_k  - B^2 \delta _{ik} /2} \right)/4\pi $
are  entered. We will be interested only in small oscillations in
the  given system. It allows producing linearization on small
amplitude deviations from the equilibrium values. Then
 \begin{equation}\label{(6)}
\pi _{ik}  = \delta _{ik} \left( {\left( {\partial P/\partial \rho
}
 \right)_s \rho  + \left( {\partial P/\partial s} \right)_\rho  s} \right)
 - \left( {\left\langle {B_i B_k } \right\rangle  - \left\langle {B_l B_l }
  \right\rangle \delta _{ik} /2} \right)/4\pi,
\end{equation}
  where deviations of correlation moments of the field from their
equilibrium values is considered also having the first order of
smallness. Apparently, Maxwell tensor of tensions enters in
\ref{(5)}
 which in the MHD approximation
is fully determined by the second moment $\left\langle {B_i B_k }
\right\rangle $. Time equation for the indicated moment we will
receives from \ref{(4)}, multiplying on $B_k $
 in that spatio-time point and making symmetrization. After averaging on accidental phases
 (\cite{[6]}  p. 439)
 and linearization we have equation
\begin{equation}\label{(7)}
\partial _t \left\langle {B_i B_k } \right\rangle  = \eta
_{iklmnp} \left\langle {B_l B_m } \right\rangle _0 \partial _n
v_p,
\end{equation}                    where tensor $\eta _{iklmnp}  =
\delta _{ip} \delta _{km} \delta _{ln}  + \delta _{im} \delta
_{kp} \delta _{ln}  - \delta _{il} \delta _{km} \delta _{np}  -
\delta _{im} \delta _{kl} \delta _{np} $ is entered. We will
neglect by thermal fluctuations. In tensor of momentum  stream
\ref{(6)} enters deviation of entropy. For simplicity we will take
interest in adiabatic processes, then \begin{equation}\label{(8)}
\partial _t v_i  + \partial _k \left\{ {\delta _{ik} v_s^2 \rho -
\left( {\left\langle {B_i B_k } \right\rangle - \left\langle {B_l
B_l } \right\rangle \delta _{ik} /2} \right)/4\pi } \right\}/\rho
_0  = 0.
\end{equation}            Here $\rho _0 $
is equilibrium value of mass density, $v_s^2  = \left( {\partial
P/\partial \rho } \right)_s $ is adiabatic velocity of sound in a
nonmagnetized liquid. Also make linearization of Eq. \ref{(1)}
\begin{equation}\label{(9)}
\partial _t \rho  + \rho _0 \partial _i v_i  = 0.
\end{equation}  System
of equations \ref{(9)}, \ref{(8)} and \ref{(7)} for variables
$\rho $ , $v_i $
 and $\left\langle {B_i B_k } \right\rangle $ is
 closed.

\section{Adiabatic one-dimension waves of small amplitude}

We will consider one-dimension waves, we will directs coordinate
axis $x_3 $ along direction of distribution. Lets all MHD values
depend only of $x_3 $
 and $t$. Let the constant magnetic field have isotropic centered
second moment, and also, selected direction for first moment.
According to the done suppositions about statistics of the
magnetic field  the equilibrium value of correlation moment is\[
\left\langle {B_l B_m } \right\rangle _0  = \left\langle {B_0^2 }
\right\rangle \delta _{lm} /3 + B_{0l} B_{0m} = const.\] Without
limitation of generality we can choose an axis $\vec x_1  \bot
\vec B_0 $,

i.e. $\vec B_0  = \left( {0,B_0 \sin \theta,B_0 \cos \theta }
\right)$, where $\theta $
  is angle between the first moment of the constant field and the wave. Equation
  \ref{(7)} is symmetric on tensor indexes $i$
and $k$, that is why contains 6 equations for component of
symmetric tensor $\left\langle {B_i B_k } \right\rangle $. It is
comfortably to represent ten equations of the system \ref{(7)} -
\ref{(9)} in a matrix form
\begin{equation}\label{(10)} \partial _t \Psi _\alpha + {\rm
Z}_{\alpha \beta }
\partial _3 \Psi _\beta   = 0.  \end{equation}                      Here
the vector of state \begin{equation}\label{(11)} \Psi  =
 \left( {\rho,v_1,v_2,v_3
,\left\langle {B_1 B_1 } \right\rangle,\left\langle {B_1 B_2 }
\right\rangle,\left\langle {B_1 B_3 } \right\rangle,\left\langle
{B_2 B_2 } \right\rangle,\left\langle {B_2 B_3 } \right\rangle
,\left\langle {B_3 B_3 } \right\rangle } \right)
\end{equation}
and matrix with next nonzero components \[{\rm Z}_{14}  = \rho
_0,\, {\rm Z}_{27}  = Z_{39} =  - 2Z_{45} =  - 2Z_{48} = 2Z_{410}
= - 1/4\pi \rho _0,\, Z_{41} = v_s^2 /\rho _0,\] \[Z_{54}=
2\left\langle {B_0^2 } \right\rangle /3,\, Z_{84}  = 2\left\langle
{B_0^2 } \right\rangle /3 + 2B 0^2 \sin ^2 \theta,\] \[  Z_{62} =
- 2Z_{83} = Z_{94} = 2\left\langle {B_0^2 } \right\rangle /3 + 2B
0^2 \sin \theta \cos \theta,\]
\begin{equation}\label{(12)}
Z_{72} = Z_{93} =  - \left\langle {B_0^2 } \right\rangle /3 -
B_0^2 \cos ^2 \theta
\end{equation}
are entered. In a plane one dimension wave dependence of the state
vector on a coordinate and time looks like (\cite{[2]}  p.49-55)
\begin{equation}\label{(13)} \Psi _\alpha = A_\alpha  \exp \left(
{ikx_3 - i\omega t} \right).
\end{equation}
Substitution Eq. \ref{(13)} in Eq. \ref{(10)} gives
\begin{equation}\label{(14)} {\rm Z}_{\alpha \beta } A_\beta =
VA_\alpha,
\end{equation}  where $V =
\omega /k$
  is phase velocity of wave, $A_\alpha  $
is right eigenvector of matrix ${\rm Z}$. It is comfortably to
enter denotations for Alfven velocity $v_A  = B_0 /\sqrt {4\pi
\rho _0 } $
 \cite{[2]}, and also for similar on a form velocity arising up due to the centered moment different from a zero
 $
v_t  = \sqrt {\left\langle {B_0^2 } \right\rangle /12\pi \rho } $
 \cite{[3]}. Solving Eq. \ref{(14)} by the standard way  we find the eigenvalues $V$
of matrix ${\rm Z}$

\[V = \left\{ {0,0,0,0, - \sqrt {v_t^2  + v_A^2 \cos ^2 \theta }
,\sqrt {v_t^2  + v_A^2 \cos ^2 \theta },} \right.\] \[- \sqrt
{\left( {3v _t^2  + v_s^2  + v_A^2 } \right) - \sqrt {\left(
{v_A^2 - v_t^2  - v_s^2 } \right)^2  - 4v _A^2 \left( {v_t^2  +
v_s^2 } \right)\cos ^2 \theta } } /\sqrt 2,\] \[\sqrt {\left( {3v
_t^2  + v_s^2  + v_A^2 } \right) - \sqrt {\left( {v_A^2  - v_t^2 -
v_s^2 } \right)^2  - 4v _A^2 \left( {v_t^2  + v_s^2 } \right)\cos
^2 \theta } } /\sqrt 2,\]               \[ - \sqrt {\left( {3v
_t^2 + v_s^2  + v_A^2 } \right) + \sqrt {\left( {v_A^2  - v_t^2  -
v_s^2 } \right)^2  - 4v _A^2 \left( {v_t^2  + v_s^2 } \right)\cos
^2 \theta } } /\sqrt 2,\]
\begin{equation}\label{(15)}
 \left. {\sqrt {\left(
{3v _t^2  + v_s^2  + v_A^2 } \right) + \sqrt {\left( {v_A^2  -
v_t^2  - v_s^2 } \right)^2  - 4v _A^2 \left( {v_t^2  + v_s^2 }
\right)\cos ^2 \theta } } /\sqrt 2 } \right\},
\end{equation}         and
also eigenvectors, which have bulky expressions.
 Unspreading perturbations of one of diagonal
 elements of tensor of the field second moment and density, and also perturbation of
 component $\left\langle {B_1 B_2 } \right\rangle $  correspond to the values $V = 0$. Six spreading waves correspond to interacting
 the MHD and new sound waves. Fifth and sixth
eigenvalues correspond to  the mode of transversal velocity
oscillations along an axis $x_1 $
 (perpendicular to both the wave and field $
\vec B_0 $ ) and components of correlation moment $\left\langle
{B_1 B_2 } \right\rangle $
 and $\left\langle {B_1 B_3 } \right\rangle $. Oscillations of velocity  along axes $x_2 $
 and $x_3 $, density of mass, components $\left\langle {B_2 B_3 }
\right\rangle $
 and diagonal components $\left\langle {B_1 B_1 } \right\rangle $
 and $\left\langle {B_2 B_2 } \right\rangle $ correspond to   the other eigenvalues.
    From the received solution in supposition of absence of correlations of the magnetic field
    $\left\langle {B_0^2 } \right\rangle  = 0$
 the standard MHD solution  as Alfven and two magnetosound waves \cite{[2]},\cite{[6]} follows.
    If to assume that the constant magnetic field is isotropic
    $
B_0  = 0 $, we have the following solution: \[V = \left\{
{0,0,0,0, - v_t,v_t, - v_t,v_t, - \sqrt {v_s^2  + 2v_t^2 },\sqrt
{v_s^2 + 2v_t^2 } } \right\}.\] Fifth and sixth eigenvalues
correspond to    the mode of transversal oscillations of velocity
along an axis $x_1 $
 and components $\left\langle {B_1 B_3 } \right\rangle $. The seventh and eighth eigenvalues correspond
 to  the mode of transversal
oscillations of velocity along an axis $x_2 $
 and components $\left\langle {B_2 B_3 } \right\rangle $. The last two eigenvalues correspond to
  the mode of longitudinal
oscillations of velocity  along an axis $x_3 $, densities of mass
and diagonal components $\left\langle {B_1 B_1 } \right\rangle $
 and $\left\langle {B_2 B_2 } \right\rangle $. That in relation to mass velocity there are two transversal
modes of oscillations with velocity $ v_t$
 and one longitudinal with velocity $
v_l  = \sqrt {v_s^2  + 2v_t^2 } $. There is a situation in this
sense fully similar to the sound modes in an isotropic solid
(\cite{[3]},\cite{[7]} p. 124-128) and the centered second moment
of magnetic field determines the "module of displacement"  $\mu  =
\left\langle {B_0^2 } \right\rangle /12\pi $.


\section{Conclusions}
\begin{itemize}
\item  Evolution of magnetized conducting liquid (or plasma) with
 the accidental field in approximation of ideal hydrodynamics is studied. Magnetic
 field as accidental value is described by  value of
second moment of magnetic induction. The linear system of
equations for the density of mass, velocity and tensor of the
second moment magnetic induction is received which allowed to
study adiabatic modes in this system.

\item Three modes of hydrodynamic oscillations are found in supposition
of presence of constant first and isotropic centered second
moments of the magnetic field. These modes turn into the standard
MHD modes in absence of constant centered second moment. In
default of the constant first moment two transversal with
coincident phase velocities and longitudinal sound modes are
received.

\end{itemize}
This work was supported by the State Foundation for Fundamental
Research of Ukraine (project No.25.2/102).



\end{document}